\documentclass[11pt,twoside]{article}
\usepackage{./asp2014}

\resetcounters

\begin{document}

\title{[C{\sc ii}] 158$\mu$m Emission from $z \ge 10$ Galaxies}

\author{C. L. Carilli,$^{1,2}$ E.J. Murphy,$^3$ A. Ferrara,$^4$ and P. Dayal$^5$}
\affil{$^1$National Radio Astronomy Observatory, Socorro, NM 87801; \email{ccarilli@nrao.edu}}
\affil{$^2$Cavendish Astrophysics, Cambridge, UK}
\affil{$^3$National Radio Astronomy Observatory, Charlottesville, VA 22903; \email{emurphy@nrao.edu}}
\affil{$^4$Scuola Normale Superiore, Piazza dei Cavalieri 7, I-56126 Pisa, Italy}
\affil{$^5$Kapteyn Astronomical Institute, University of Groningen, 
P.O. Box 800, 9700 AV Groningen, The Netherlands}




\begin{abstract}
We consider the capabilities of ALMA and the ngVLA 
to detect and image the [C{\sc ii}]
158\,$\mu$m line from galaxies into the cosmic `dark ages' ($z \sim
10$ to 20). The [C{\sc ii}] line may prove to be a powerful tool in
determining spectroscopic redshifts, and galaxy dynamics, for the
first galaxies.  In 40\,hr, ALMA has the sensitivity to detect the
integrated [C{\sc ii}] line emission from a moderate metallicity, active
star-forming galaxy [$Z_A = 0.2\,Z_{\odot}$; star formation rate (SFR)
= 5\,$M_\odot$\,yr$^{-1}$], at $z = 10$ at a significance of
6$\sigma$. The ngVLA will detect
the integrated [C{\sc ii}] line emission from a Milky-Way like star
formation rate galaxy ($Z_{A} = 0.2\,Z_{\odot}$, SFR =
1\,$M_\odot$\,yr$^{-1}$), at $z = 15$ at a significance of
6$\sigma$. Imaging simulations show that the ngVLA can determine
rotation dynamics for active star-forming galaxies at $z \sim 15$, if
they exist. The [C{\sc ii}] detection rate in blind surveys will be slow (of
order unity per 40\,hr pointing).\footnote{This paper
is a brief synopsis of a paper presented in the Astrophysical Journal 
\citep{carilli17}. We refer the interested reader to the Journal 
article for more detail.}

\end{abstract}


\section{Introduction}

The $z \sim 15$ Universe is at the edge of our current understanding.
A handful of theoretical studies have speculated on the cosmic star
formation rate (SFR) density at these redshifts \citep{mashian16,
duffy17, cp10, dayal14, yue15, topping15}.  Observational constraints
on extreme redshift galaxies are poor, based on extrapolation of the
few galaxies and AGN known at $z \sim 7$ to 8, and the even fewer
galaxy candidates at $z \sim 8$ to 11.

Encouraging results come from observations of a relatively mature
interstellar medium, and active star formation, in some of the very
high redshift sources discovered to date.  The last few years have
seen an explosion in the number of [C{\sc ii}] 158$\mu$m detections at high redshift,
including high resolution imaging of the gas dynamics on kpc-scales in
both AGN host galaxies and in more normal star-forming galaxies at $z
\sim 5.5$ to 7.5 \citep[see][]{carilli17} for a summary). The most 
recent results include the detection of the [O{\sc iii}]  
88$\mu$m fine structure line and/or the [C{\sc ii}] line, from galaxies at $z =
7.2$, 8.4, and 9.1 \citep{lapo17, has18a, has18b}
and the detection of strong [C{\sc ii}] and 
dust continuum emission from a quasar host galaxy at $z = 7.5$
\citep{ven17}. While encouraging, observations remain sparse,
and the most basic questions remain on the nature, and even
existence, of galaxies at $z \sim 15$.

Given the uncertainty in our knowledge of galaxies at extreme
redshifts, in this study we focus on a few simple of questions: if
such extreme redshift galaxies exist, what kind of facility is
required to detect, and possibly image, the [C{\sc ii}] 158\,$\mu$m line
emission?  How do the prospects depend on basic galaxy properties,
such as metallicity and star formation rate? And based on what little we
know of galaxy demographics at very early epochs, what kind of numbers
can we expect in blind cosmological spectral deep fields?

\section{The [C{\sc ii}] 158$\mu$m line}

The [C{\sc ii}] 158\,$\mu$m line is one of the brightest  spectral
line from star-forming galaxies at far-infrared wavelengths and
longer, carrying between 0.1\% to 1\% of the total far infrared
luminosity of star forming galaxies \citep{stac91, cw13}. The [C{\sc ii}] fine
structure line traces both neutral and ionized gas in galaxies, and is
the dominant coolant of star-forming gas in galaxies \citep{velusamy15}.  
While the line is only visible from space in
the nearby Universe, it becomes easier to observe with increasing
redshift, moving into the most sensitive bands of large ground based
millimeter telescopes, such as NOEMA\footnote{\url
{http://iram-institute.org/EN/noema-project.php}}, and the
ALMA\footnote{\url{http://www.almaobservatory.org}}.

As a predictor for the [C{\sc ii}] 158\,$\mu$m luminosity from early
galaxies we use the \citet{vallini15} relationship (their Equation
8). This theoretical and observational analysis considers in detail
the relationships between star formation rate, galaxy metallicity, and
[C{\sc ii}] luminosity. We adopt a few representative galaxy
characteristics, including the main parameters of star formation
rate, metallicity, redshift, and [C{\sc ii}] luminosity, and compare these
to the capabilities of the given facilities.

\section{Telescopes}
\label{sec:tel}

The relevant ALMA bands are 3, 4, and 5, corresponding to
frequencies of $84-116$\,GHz, $125-163$\,GHz, and $163-211$\,GHz,
respectively. These bands then cover the [C{\sc ii}] line (1900.54\,GHz rest
frequency), between $z = 10$ and 20, almost continuously, with gaps of
a few MHz due to atmospheric O$_2$ absorption at $118\,$GHz and
183\,GHz.  The current bandwidth for ALMA is 8 GHz, with a upgrade
to 16\,GHz or 32\,GHz being considered sometime in the future.  We employ the
ALMA sensitivity calculator, under good weather conditions (3rd
octile), with 50 antennas. For the sake of illustration, we adopt a
fiducial line width of 100 km s$^{-1}$ (see below) and an on-source
integration time of 40\,hr.  The rms sensitivity per channel is 21
$\mu$Jy beam$^{-1}$ channel$^{-1}$, roughly independent of frequency
due to the increasing channel width in Hz for a fixed velocity
resolution, offsetting decreasing system sensitivity with increasing frequency.

\begin{table}[h]
\caption{Facilities}
\label{tbl:fac}
\begin{center}
\begin{tabular}{lcccc}
\hline
Telescope & Redshifts & Frequencies & rms & Bandwidth \\
~ & ~ & (GHz) & ($\mu$Jy\,beam$^{-1}$) & (GHz) \\
\hline
ngVLA & $15 - 20$ & $116 - 90$ & 2.0 & 26 \\
ALMA & $10 - 15$ &  $173 - 116$ & 21 & 8 (32) \\
\hline
\vspace{0.1cm}
\end{tabular}
\end{center}
\end{table}

For the ngVLA we employ the ``Southwest" configuration, and we adopt
sensitivity parameters consistent with ngVLA memo 17 \citep{sm17}.
The maximum redshift we consider is $z = 20$, so we only consider
frequencies between 90 and 116\,GHz. The current reference design
has a nominal maximum bandwidth of 20\,GHz, although broader bandwidths
are under investigation. For the sake of number counts, the $z = 18.8$ to 20
range (90 GHz to 96 GHz), contributes very little to the total number of
sources detected in blind searches.  For the purpose of estimating the
sensitivity of the ngVLA for realistic observations, and to explore
the imaging capabilities in the event of the discovery of any
relatively luminous sources, we have employed the CASA simulation
tools \citep{cs17}, developed for the ngVLA project.

We simulate a 40\,hr observation, and we employ the CLEAN algorithm
with Briggs weighting.  We adjust the {\sc robust} parameter, the
$(u,v)$-taper, and the cell size, to give a reasonable synthesized
beam and noise performance. Our target resolution is $\sim 0\farcs4$
for detection, and $\sim 0\farcs2$ for imaging.

We adopt as a spatial and dynamical template, the observed CO 1-0
emission from the nearby star-forming disk galaxy, M\,51
\citep{helfer03}.  We arbitrarily reduce the physical size of the disk
by a factor three, with the idea that very early galaxies are likely
smaller than nearby galaxies.  Again, this exercise is for
illustrative purposes, and the input model is just a representative
spatial/dynamical template for a disk galaxy, with the relevant
parameters being size, velocity, and luminosity.  We then adjust the
line luminosity per channel per beam, to achieve a given integrated
[C{\sc ii}] 158\,$\mu$m luminosity at a given redshift.

\begin{figure}[t]
\begin{center}
\includegraphics[scale=0.6]{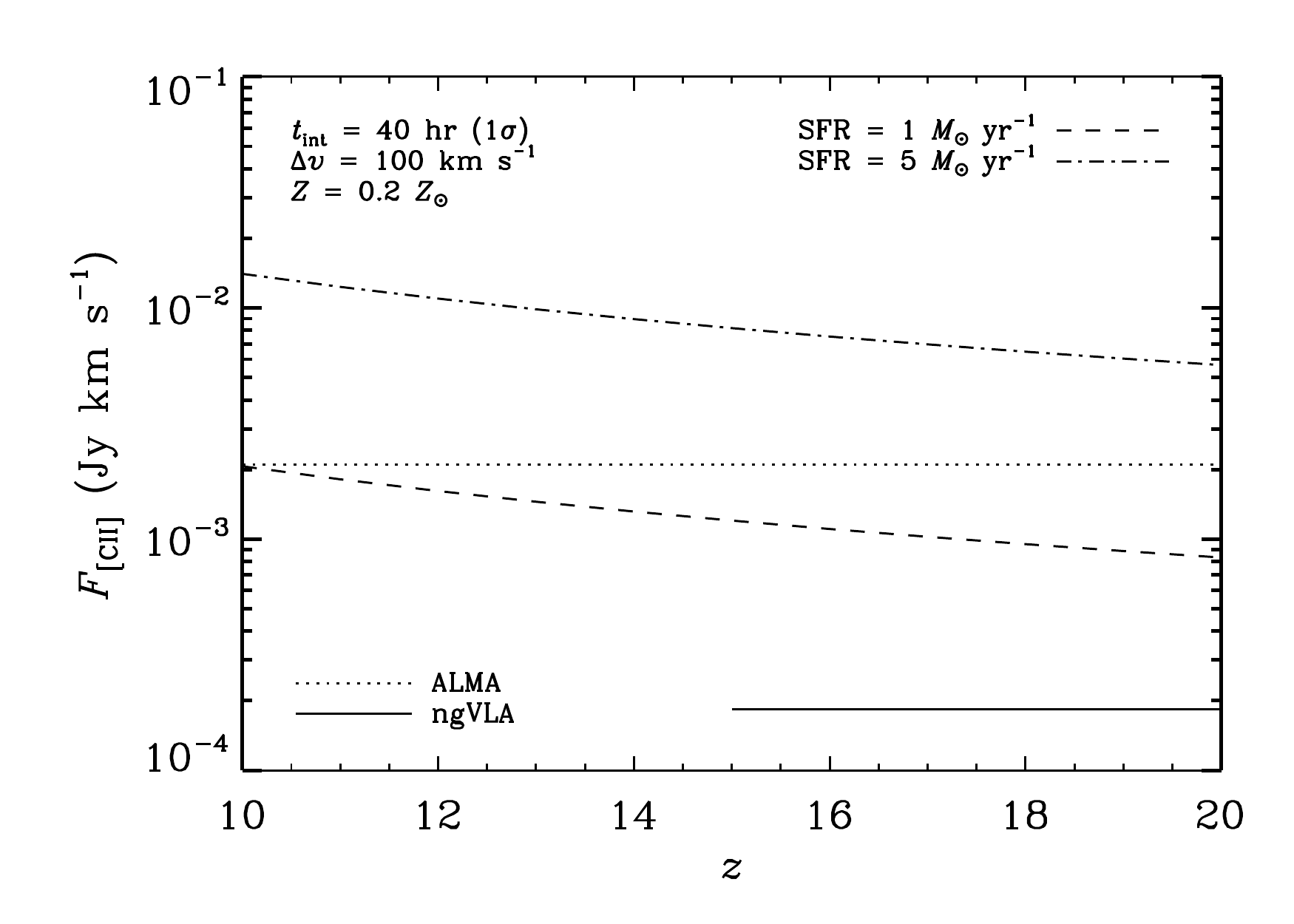}
\end{center}
\caption{
[C{\sc ii}] 158\,$\mu$m velocity integrated line flux versus redshift
for galaxies with star formation rates of 1\,$M_\odot$\,yr$^{-1}$
and 5\,$M_\odot$\,yr$^{-1}$, and metallicity of 0.2\,$Z_{\odot}$,
based on the relationship given in Equation 12 
of \citet{vallini15}. 
The rms sensitivity in a 100\,km\,s$^{-1}$ channel and 40\,hr integration
is shown for both ALMA and the ngVLA.
}
\label{fig:scii}
\end{figure}

\section{Results}

\subsection{Spectroscopic Confirmation of $z\gtrsim10$ Candidates}
\label{sec:specconf}

An obvious application of the [C{\sc ii}] 158\,$\mu$m line search will be to
determine spectroscopic redshifts for near-IR dropout candidate
galaxies at $z \sim 10$ to 20.
We start with the relationship between the [C{\sc ii}] velocity integrated
line flux, in the standard flux units of Jy km s$^{-1}$, versus
redshift. We adopt a metallically of $Z_A = 0.2\,Z_{\odot}$, and star
formation rates of 1\,$M_\odot$\,yr$^{-1}$ and
5\,$M_\odot$\,yr$^{-1}$. Figure~\ref{fig:scii} shows the predicted
[C{\sc ii}] line flux versus redshift for the two models, along with the
1$\sigma$ sensitivity of ALMA and the ngVLA. 

This image simulation shows that, in 40\,hr, the ngVLA will be
able to detect the integrated [C{\sc ii}] line emission from moderate
metallicity and star formation rate galaxies ($Z_{A} = 0.2$, SFR =
1\,$M_\odot$\,yr$^{-1}$), at $z = 15$ at a significance of
6$\sigma$. This significance reduces to 4$\sigma$ at $z= 20$.

In 40\,hr, ALMA will be able to detect the integrated [C{\sc ii}] line
emission from a higher star formation rate galaxy ($Z_A =
0.2\,Z_{\odot}$, SFR = 5\,$M_\odot$\,yr$^{-1}$), at $z = 10$ at a
significance of 6$\sigma$. This significance reduces to 4$\sigma$ at
$z= 15$. ALMA will be hard-pressed to detect a moderate metallicity
($Z_A = 0.2\,Z_{\odot}$), lower star formation rate
(1\,$M_\odot$\,yr$^{-1}$) galaxy, requiring 1000\,hr for a 5$\sigma$
detection of the velocity integrated line flux, even at $z = 10$.

\begin{figure}[h]
\begin{center}
\includegraphics[scale=0.4]{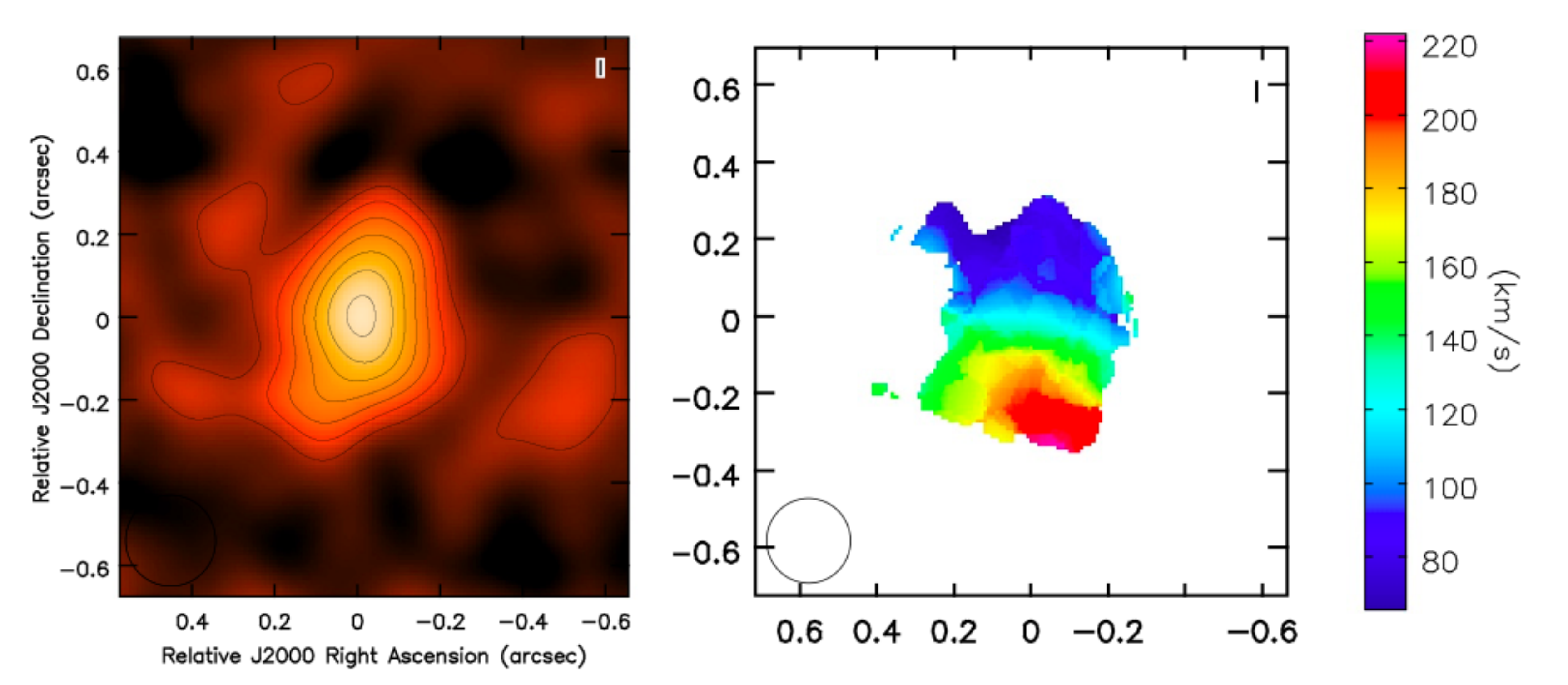}
\end{center}
\caption{
({\it Left:}) A simulated image of the velocity integrated [C{\sc ii}]
158\,$\mu$m emission from a $z = 15$ galaxy with a star formation rate
of 5\,$M_\odot$\,yr$^{-1}$, and a metallicity of 0.2\,$Z_{\odot}$,
assuming for a 40\,hr observation with the ngVLA. 
The contour levels are -6, -3, 3, 6, 9, 12, 15, 18, 21\,$\mu$Jy\,beam$^{-1}$. The rms noise on the
image is about 1.8\,$\mu$Jy\,beam$^{-1}$, and the synthesized beam
FWHM is $0\farcs22$.  ({\it Right:}) The intensity weighted mean [C{\sc ii}]
velocity (moment 1).
}
\label{fig:5M}
\end{figure}

\begin{figure}[t]
\begin{center}
\includegraphics[scale=0.6]{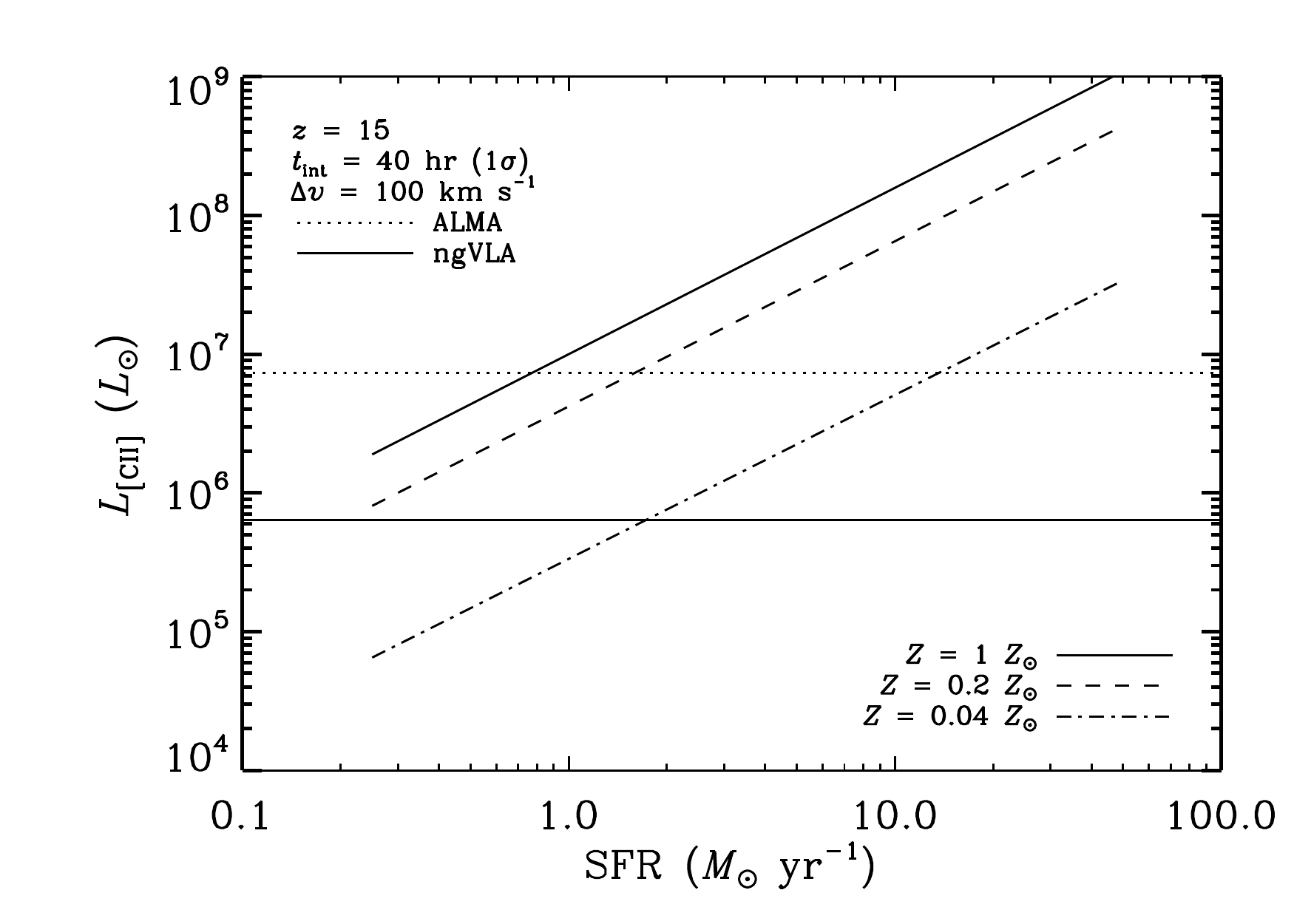}
\end{center}
\caption{[C{\sc ii}] 158\,$\mu$m line luminosity 
versus star formation 
rate and metallicity, based on the relationship given in Equation 12 
of \citet{vallini15}. Three different metallicities are shown.
Also shown is the rms sensitivity of ALMA and the ngVLA for
a galaxy at $z =  15$, assuming a 100\,km\,s$^{-1}$ 
channel and  40\,hr integration.
}
\label{fig:lcii}
\end{figure}

We next consider dependence on metallicity.  Figure~\ref{fig:lcii}
shows the relationship between [C{\sc ii}] luminosity (in Solar units), to
star formation rate, for three different metallicities: $Z_A = 0.04$,
0.2, and 1.0\,$Z_{\odot}$, for a galaxy at $z = 15$. Again shown
are the ALMA and ngVLA sensitivities in 40\,hr, 100 km s$^{-1}$
channels. The \citet{vallini15} model has the [C{\sc ii}] luminosity as a strong
function of metallicity.  If the gas has Solar metallicity, the ALMA
detection threshold ($4\sigma$) reduces to a galaxy with a star
formation rate of 2.5\,$M_\odot$\,yr$^{-1}$ (compared to
5\,$M_\odot$\,yr$^{-1}$ for $Z_A = 0.2$), while that for the ngVLA
reduces to 0.4\,$M_\odot$\,yr$^{-1}$ (compared to
1\,$M_\odot$\,yr$^{-1}$ for $Z_A = 0.2$). Conversely, for a low
metallicity galaxy of $Z_A = 0.04\,Z_{\odot}$, these values increase
to 100\,$M_\odot$\,yr$^{-1}$ and 10\,$M_\odot$\,yr$^{-1}$,
respectively.

\subsection{Kinematics of $z\gtrsim10$ Galaxies }

We investigate the potential for obtaining kinematic information from
such galaxies using the ngVLA.  We find that the best even the ngVLA
can do for a $z  = 15$, $Z_A = 0.2\,Z_{\odot}$, and SFR =
1\,$M_\odot$\,yr$^{-1}$ galaxy, in 40\,hr is a 
$5.5\sigma$ detection of the integrated emission, with little
or no dynamical information.

For a higher SFR galaxy ($z = 15$, $Z_A = 0.2\,Z_{\odot}$,
5\,$M_\odot$\,yr$^{-1}$ galaxy), in a 40\,hr observation, the ngVLA can
recover the overall rotational dynamics of the system.  The imaging
results for the velocity integrated emission (mom 0), and the
intensity weighted mean velocity (mom 1), are shown in
Figure~\ref{fig:5M}.  The beam size in this case is FWHM $\sim
0\farcs2$, and the channel images at 20\,km s$^{-1}$\,channel$^{-1}$
have an rms noise of 4.5\,$\mu$Jy beam$^{-1}$.

\subsection{The Potential for Blind Searches of $z\gtrsim10$ Galaxies}

Another application for the [C{\sc ii}] line will be blind cosmological deep
fields. The advent of very wide bandwidth spectrometers has led to a
new type of cosmological deep field, namely, spectral volumetric deep
fields, in which a three dimensional search for spectral lines can be
made, with redshift as the third dimension \citep{walter16}.
 
\begin{table}[tbh]
\footnotesize
\caption{Number of Detections per 40\,hr Pointing}
\label{tbl:detrate}
\begin{tabular}{lcccc}\hline
Model & ngVLA & ngVLA & ALMA & ALMA  \\
  & $15 <z< 16$ & $15 <z< 20$ & $10<z< 10.5$ (8\,GHz) &  $11<z< 14$ (32\,GHz) \\
\hline
CP10, 1\,$M_\odot$\,yr$^{-1}$ & 0.29 & 1.3 & -- & -- \\
CP10, 5\,$M_\odot$\,yr$^{-1}$ & 0.11 & 0.48 & 0.29 & 0.68 \\
Dayal14, 1\,$M_\odot$\,yr$^{-1}$ & 0.36 & 0.64 & -- & -- \\
Dayal14, 5\,$M_\odot$\,yr$^{-1}$ & $6.9\times 10^{-4}$ & $7.3\times 10^{-4}$ & 2.8 & 1.4 \\
\hline
\vspace{0.1cm}
\end{tabular}
\end{table}

Given the large uncertainty in the predicted galaxy luminosity
function beyond $z \sim 10$, we investigated two theoretical
predictions for the [C{\sc ii}] luminosity function with very different
methodologies.  The first method used star forming galaxy number
counts of \citet[][CP10]{cp10}.  These galaxy counts are based on
backward-evolving models for the infrared luminosity function of
\citet{ce01}, anchored by a variety of observational data from 
{\it Spitzer} and  {\it Herschel}.

The second method employed the calculations of high redshift galaxy
formation of \citet{dayal14}.  This model aims at isolating the
essential physics driving early galaxy formation via a merger-tree
based semi-analytical model including the key physics of star
formation, supernova feedback, and the growth of progressively more
massive systems (via halo mergers and gas accretion). This model
reproduces well both the slope and amplitude of the UV LF from $z=5$
to $z=10$.

The two models predict the cummulative co-moving number density of
star-forming galaxies above a given star formation rate as a function
of redshift. These values can be converted to cummulative number
densities of [C{\sc ii}] emitting galaxies as a function of line flux, using
the models of \citet{vallini15}.  The [C{\sc ii}] number densities vs. flux
can then be turned into the number of observed galaxies in a given
integration time, bandwidth, and field of view, using the
sensitivities, field sizes, and bandwidths of the ngVLA and ALMA, as
discussed in \S\ref{sec:tel}.

The ngVLA covers the $90-116$\,GHz range, 
corresponding to $z = 20$ to 15. We also consider just
the number of galaxies between $z = 15$ and 16.  ALMA has receivers
that will cover from $z = 10$ to 15, or frequencies from 173\,GHz to
116\,GHz. Currently, the bandwidth is limited to 8\,GHz. We consider an
8\,GHz blind search in the Band 5 from 165\,GHz to 173\,GHz ($z =
10.5$ to 10), and one covering most of Band 4 with a hypothetical
32\,GHz bandwidth system, from 126\,GHz to 158\,GHz ($z =11$ to 14).

In Table \ref{tbl:detrate}, we tabulate the number of galaxies
detected in [C{\sc ii}] emission per 40\,hr integration per frequency
tuning, for the ngVLA and ALMA, and for the different models. For the
ngVLA, and for SFR $\ge 1$\,$M_\odot$\,yr$^{-1}$, the models predict
that one to two independent pointings will be required to
detect one galaxy over the full redshift range, on average. For the
CP10 model, these sources have a broader redshift distribution, with
22\% of the sources at $z =15$ to 16. For the Dayal14 model, the
majority (64\%), of the sources are in this lowest redshift bin.

For ALMA and an SFR $\ge 5$\,$M_\odot$\,yr$^{-1}$, the predicted number
of detections differs significantly between models. For the 8\,GHz
bandwidth search in Band 5 ($z = 10$ to 10.5), the CP10 model requires
about three pointings for a single detection, on average, while the
Dayal14 model has more low redshift, brighter galaxies, with three
sources per pointing expected. For the hypothetical 32\,GHz bandwidth
search in Band 4 ($z = 11$ to 14), the values are roughly two
pointings needed for a single detection for the CP10 model, and one
pointing needed for the Dayal14 model.

Overall, the detection rates in blind surveys will be slow (of order
unity per 40\,hr pointing). However, the observations are well suited
to commensal searches on all programs employing the very wide bands
that may be available in future. 

A key issue in blind searches is spurious detections and verifying
sources, especially given the large number of voxels searched for
emission in the proposed surveys \citep[see][]{carilli17}).  Recent
blind line searches have developed some techniques for making
statistical corrections to number counts based on e.g., comparing the
number of negative and positive detections at a give level
\citep{decarli16, walter16, aravena16}. However, the problem still
remains as to how to verify that a given detection is associated with
a $z > 10$ galaxy.  One possible method will be broad band near-IR
colors from e,g., \textit{JWST}, or large ground based telescopes.
Likewise, follow-up spectroscopy with large ground and 
space-based telescopes  may reveal atomic
lines \citep{barrow17}. Lastly, ALMA could be used to search for
[O{\sc iii}] 88$\mu$m emission, in cases of low metallicity galaxies
\citep{cormier15, has18a, has18b}.

\section{Conclusions}

We have considered observing [C{\sc ii}] 158$\mu$m emission from $z = 10$ to
20 galaxies. The [C{\sc ii}] line may prove to be a powerful tool to determine
spectroscopic redshifts, and galaxy dynamics, for the first galaxies
at the end of the dark ages, such as identified as near-IR dropout
candidates by \textit{JWST}. 

In 40\,hr, the ngVLA has the sensitivity to detect the integrated [C{\sc ii}] line
emission from moderate metallicity and (Milky-Way like) star formation
rate galaxies ($Z_{A} = 0.2$, SFR = 1\,$M_\odot$\,yr$^{-1}$), at $z =
15$ at a significance of 6$\sigma$. This significance reduces to
4$\sigma$ at $z= 20$. In 40\,hr, ALMA has the sensitivity to detect the
integrated [C{\sc ii}] line emission from a higher star formation rate
galaxy ($Z_A = 0.2\,Z_{\odot}$, SFR = 5\,$M_\odot$\,yr$^{-1}$), at $z
= 10$ at a significance of 6$\sigma$. This significance reduces to
4$\sigma$ at $z= 15$.  We also consider dependencies on metallically
and star formation rate. 

We perform imaging simulations using a plausible model for the gas
dynamics of disk galaxies, scaled to the sizes and luminosities
expected for these early galaxies. The ngVLA will recover rotation
dynamics for active star-forming galaxies ($\gtrsim
5\,M_\odot$\,yr$^{-1}$ at $z \sim 15$), in reasonable integration
times.

We adopt two models for very high redshift galaxy formation, and
calculate the expected detection rate for [C{\sc ii}] emission at $z \sim
10$ to 20, in blind, wide bandwidth, spectroscopic deep fields. The
detection rates in blind surveys will be slow (of order unity per
40\,hr pointing). However, the observations are well suited to
commensal searches on all programs employing the very wide bands that
may be available in future. 

\vskip 0.2in 

\acknowledgements PD acknowledges support from the European Research
Council's starting grant ERC StG-717001 and from the European
Commission's and University of Groningen's CO-FUND Rosalind Franklin
program. The National Radio Astronomy Observatory is a facilites of
the National Science Foundation operated under cooperative agreement
by Associated Universities, Inc.. We thank Ranga-Ram Chary for
discussions on the models and the paper, and N. Scoville for bringing
up the possibility of very high $z$ [C{\sc ii}] in the ngVLA context.



\end{document}